\documentclass[aps,pre,showpacs,twocolumn]{revtex4}
\usepackage{bm}
\usepackage{graphicx}
\bibstyle{approve.bib}
\usepackage{amssymb}
\usepackage{amsmath}
\usepackage{esint}
\usepackage{epstopdf}
\usepackage{color}
\usepackage{gensymb}
\usepackage{mathtools}
\usepackage{suffix}

\newcommand{\be}{\begin{equation}}
\newcommand{\ee}{\end{equation}}
\newcommand{\bea}{\begin{eqnarray}}
\newcommand{\eea}{\end{eqnarray}}
\newcommand{\lan}{\left\langle}
\newcommand{\ran}{\right\rangle}
\newcommand{\br}{\mathbf{r}}

\newcommand{\ba}{{\boldsymbol{a}}}

\newcommand{\bq}{\mathbf{q}}

\newcommand{\bu}{\mathbf{u}}

\newcommand{\bo}{\mathbf{\Omega}}
\newcommand{\bom}{\mathbf{\Omega}}
\newcommand{\bk}{\mathbf{k}}

\newcommand{\bv}{\mathbf{v}}
\newcommand{\e}{\varepsilon}
\newcommand{\te}{\tilde{\varepsilon}}

\newcommand{\tv}{\tilde{v}}

\newcommand{\tG}{\tilde{G}}

\newcommand{\ce}{_{\rm c}}

\newcommand{\n}{_{\rm n}}
\newcommand{\s}{_{\rm s}}
\newcommand{\w}{_{\rm w}}
\newcommand{\g}{_{\rm G}}

\newcommand{\hn}{\hat{\rho}}
\newcommand{\hd}{\hat{n}}

\newcommand{\ph}{\mathbf{\Phi}}
\newcommand{\h}{_{\rm h}}

\begin{document}

\title{Explicit Solvent Theory of Salt-Induced Dielectric Decrement}

\author{Sahin Buyukdagli}
\address{Department of Physics, Bilkent University, Ankara 06800, Turkey}

\begin{abstract}

We introduce a field-theoretic electrolyte model composed of structured solvent molecules and salt ions coupled by electrostatic and hard-core (HC) interactions. Within this explicit solvent theory, we characterize the salt-driven dielectric decrement beyond weak-coupling (WC) electrostatics. The WC approximation of prior formalisms is relaxed by treating the salt charges via a virial expansion. This virial approach enables the explicit inclusion of the many-body salt-solvent interactions, and directly leads to the experimentally observed linear decay of the electrolyte permittivity with added dilute salt. The permittivity formula emerging from our approach indicates that the reduction of the solvent permittivity is induced by the salt screening of the polarization charges suppressing the dielectric response of the solvent. By comparison with experiments, we also show that the salt-dressed permittivity formula can equally reproduce the attenuation of the electrolyte permittivity with rising temperature, the thermal decay of the dielectric decrement, and its intensification with the salt valency. The consistent qualitative agreement of our theory with this wide range of experimental trends points out the electrostatic ion-solvent correlations as the primary mechanism behind the salt-induced dielectric decrement.

\end{abstract}


\date{\today}

\maketitle

\section{Introduction}

The process of salt solvation in water is a vital regulator of biological processes sustaining life on earth. By preventing oppositely charged ions from forming neutral pairs, solvation acts as a key mediator for the salt-induced regulation of various mechanisms ranging from the stability of DNA molecules around histones to nanofluidic charge transport through ion channels in the cell medium~\cite{Stokes,Isr,Hille}. This universal role implies that the accurate characterization of solvation is an essential task to decipher the functioning of living organisms.

Ionic solvation originates from the dielectric screening of the mobile charges by the strongly dipolar water molecules. Within the framework of solvent-implicit electrostatics, the strength of this effect is characterized by the macroscopic dielectric permittivity setting the magnitude of the electrostatic counterion interactions. The first limitation of the implicit solvent electrostatics is the uniformity of the dielectric permittivity, which is contradicted by AFM experiments~\cite{expdiel} and MD simulations~\cite{Hans,prlnetz} indicating the spatially inhomogeneous dielectric response of polar solvents around charge sources. 

With the aim to account for the non-uniformity of the hydration by water, earlier non-local electrostatic formalisms integrated structured dielectric permittivities into the electrostatic equations of state~\cite{Kor,yar}. By generalizing the point-dipole model of Refs.~\cite{dunyuk,orland1} to solvent molecules with finite size, we introduced the first field-theoretic formulation of non-local electrostatics able to map from the intramolecular solvent structure to inhomogeneous dielectric response~\cite{PRE1,JCP2013}. In these works, we showed that the consideration of the extended charge structure of solvent molecules directly leads to the structured dielectric permittivity profiles observed in AFM experiments~\cite{expdiel} and explicit solvent simulations~\cite{Hans,prlnetz}.

An additional limitation of the solvent-implicit electrostatics is the experimentally confirmed inconsistency of the hierarchy between the salt-driven Debye screening and the solvent-induced dielectric screening. Namely, in the implicit solvent formulation of electrolyte solutions, the strength of the salt screening quantified by the Debye-H\"{u}ckel (DH) parameter depends on the dielectric permittivity of the liquid, whereas this permittivity is assumed to be independent of the salt concentration. 

The inaccuracy of this assumption has been demonstrated by the experimental observation of dielectric decrement upon salt addition into water. Over the past six decades, various characteristics of this effect have been probed by a wide range of experiments. The common observation of these experiments can be mentioned as the linear decay of the dielectric permittivity with added dilute salt and the transition to a non-linear decrement regime beyond molar salt concentrations, the reduction of the electrolyte permittivity with rising temperature, and the amplification of the dielectric decrement with the valency of the salt ions ~\cite{Hasted1948,Haggis1952,Harris1957,Hasted1958,Giese1970,Sridhar1990,Wei1992,Nortemann1997,Buchner1999,Chen2003,Loginova2006,Mollerup2015}.

Owing to the complexity of the non-local electrostatics, salt-induced dielectric decrement has been mainly investigated by implicit solvent theories and numerical simulations. Namely, the effect of the salt-dependent permittivity on the activity coefficient of NaCl solutions has been considered by Valisk\'{o} and Boda via solvent-implicit simulations~\cite{Boda2014}. By comparison with experiments, Saric at al. studied the dielectric decrement by explicit solvent simulations~\cite{Saric2020}. The problem has been equally analyzed via a microfield approach~\cite{Gavish2016}, the dressed-ion theory~\cite{Persson2017}, and the Ornstein-Zernick formalism~\cite{Friedman1982}. 

The consistent characterization of the dielectric decrement effect involving salt and solvent molecules on an equal footing requires the investigation of the problem by an explicit solvent framework. Prior solvent-explicit theories showed that the WC treatment of purely electrostatic models does not capture the salt-driven dielectric decrement. Indeed, within the framework of a point-dipole model, the calculation of the electrolyte permittivity via a one-loop-level WC approach was shown to yield the Langevin relation corresponding to the pure solvent permittivity~\cite{Andelman2013,Andelman2018}. Moreover, our gaussian-level WC analysis of the field-theoretic solvent model of Ref.~\cite{PRE1} led to an identical result~\cite{JCP2013}. Subsequently, Adar et al. showed that the incorporation of the excluded-volume effects to the liquid model of Refs.~\cite{Andelman2013,Andelman2018} generates the effect of dielectric decrement by salt addition~\cite{Andelman2018II}.

With the aim to carry out the first analysis of the salt-induced dielectric decrement beyond WC electrostatics, we introduce a field-theoretic model of explicit solvent molecules and salt charges coupled by electrostatic and HC interactions. The extension beyond the WC regime is presented in Section~\ref{fm}. Our strategy is inspired by the hybrid perturbative treatment of solvent-implicit electrolyte mixtures introduced by Kanduc at al.~\cite{Podgornik2011} and extended by us~\cite{JCP2020} to analyze correlation effects in nanofluidic charge transport~\cite{JPCB2020} and polymer translocation~\cite{Langmuir2022}. Namely, by exploiting the strong contrast between the salt and solvent concentrations, we incorporate the salt fluctuations via a virial expansion. This virial treatment limits the formalism to moderate salt concentrations but enables the explicit inclusion of the salt-solvent correlations truncated by the WC approximations of the aforementioned works. Via the solution of the electrostatic Schwinger-Dyson (SD) equation, we show that this  improvement directly gives rise to an electrolyte permittivity linearly decaying with the salt concentration. 

In Section~\ref{res}, we confront this theoretical result with experimental permittivity data and show that our formalism can equally capture the experimentally observed reduction of the electrolyte permittivity with the rise of the temperature, the thermal decay of the dielectric decrement, and its enhancement with the salt valency. We also carry out a detailed analysis of the salt effects on the spatially non-uniform macroscopic dielectric permittivity profile. We show that salt-solvent correlations reducing the magnitude and the range of the dielectric screening also suppress the non-locality of the electrostatic interactions by smoothing the dielectric permittivity structure in the vicinity of charge sources. The limitations of the model and possible extensions are elaborated in Conclusions.

\section{Explicit solvent formalism}
\label{fm}

\subsection{Derivation of the field-theoretic partition function}

\begin{figure}
\includegraphics[width=1.\linewidth]{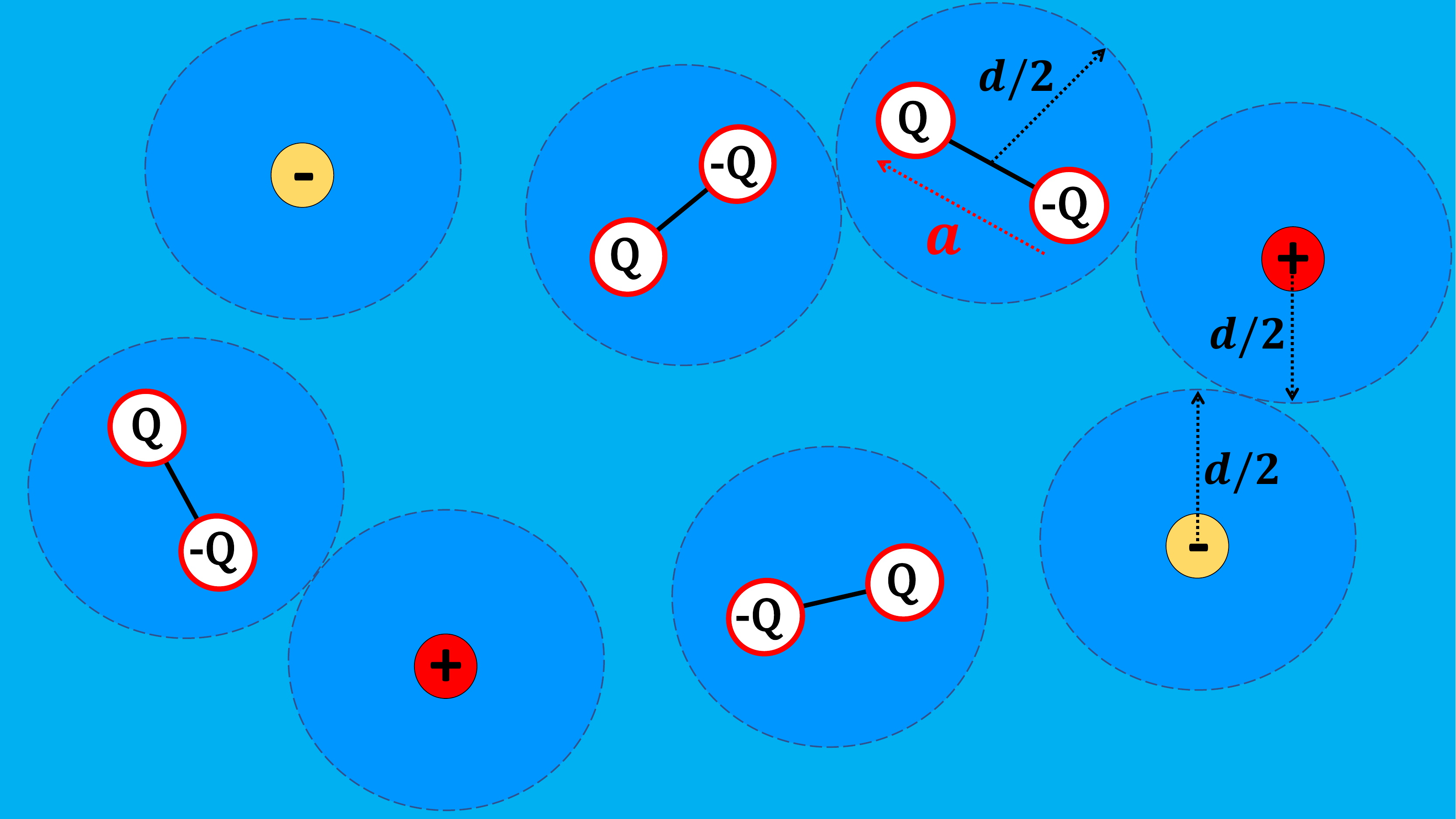}
\caption{(Color online) Schematic depiction of the charged liquid. The liquid is composed of dipolar solvent molecules with size $a=1$ {\AA} and concentration $\rho\w=55$ M, and $p$ species of monopolar salt ions. The ionic species $j$ has valency $q_j$ and concentration $\rho_j$, with $1\leq j\leq p$. Each salt ion or solvent dipole is confined to a HC sphere with diameter $d>a$ whose center coincides with the C.M. of the enclosed particle.}
\label{fig1} 
\end{figure}

We introduce here our electrolyte model depicted in Fig.~\ref{fig1}. The electrolyte is composed of dipolar solvent molecules and $p$ species of point-like salt ions. Each ion of the species $j$ has valency $q_j$, fugacity $\Lambda_j$, and concentration $\rho_j$. Moreover, the solvent dipoles of size $a$, fugacity $\Lambda\w$, and liquid water concentration $\rho\w=55$ M carry the terminal charges $\pm Q=\pm1$. 

Each solvent molecule or salt particle is placed inside an impenetrable hard-core (HC) sphere of diameter $d>a$. The center of this sphere coincides with the center of mass (C.M.) of the enclosed molecule. Thus, the pairwise HC interactions between the molecules are mediated by the potential $v\h(\br)$ defined as
\be\label{hcp}
e^{-v\h(\br)}=H(r-d),
\ee
where $H(x)$ stands for the Heaviside step function. Moreover, the electrostatic interactions between all charges are mediated by the Coulomb potential $v\ce(\br)=\ell_{\rm B}/r$ including the Bjerrum length $\ell_{\rm B}=e^2/(4\pi\e_0 k_{\rm B}T)$ in vacuum, with the electron charge $e$, the Boltzmann constant $k_{\rm B}$, the ambient temperature $T$, and the vacuum permittivity $\e_0$. In the present work, the dielectric permittivities are expressed in units of the vacuum permittivity $\e_0$.

The canonical partition function of the liquid reads
\be
\label{eq1}
Z\ce=\prod_{i=1}^{N\w}\int\frac{\mathrm{d}^2\bo_i}{4\pi}\mathrm{d}^3\bu_i\prod_{j=1}^p\prod_{k=1}^{N_j}\int\mathrm{d}^3\bv_{jk} e^{-\beta(E-\epsilon)},
\ee
with the number of solvent molecules $N\w$, the C.M. position $\bu_i$ of the dipole $i$, the number of ions $N_j$ of the species $j$, and the C.M. position $\bv_{jk}$ of the ion with number $k$ in the species $j$. Moreover, the total energy in Eq.~(\ref{eq1}) reads $E=E\ce+E\h+E\n$, where we defined the electrostatic, HC, and steric energy components as
\bea
\label{eq3}
\beta E\ce&=&\frac{1}{2}\int\mathrm{d^3}\br\mathrm{d^3}\br'\hn\ce(\br)v\ce(\br-\br')\hn\ce(\br'),\\
\label{eq4}
\beta E\h&=&\frac{1}{2}\int\mathrm{d^3}\br\mathrm{d^3}\br'\hn\n(\br)v\h(\br-\br')\hn\n(\br'),\\
\label{eq5}
\beta E\n&=&\sum_{i=1}^{N\w}w\w(\bu_i,\ba_i)+\sum_{j=1}^p\sum_{k=1}^{N_j}w_j(\bv_{jk}),
\eea
respectively. 

In Eqs.~(\ref{eq3})-(\ref{eq4}), the charge and number density operators are respectively defined as
\bea
\label{eq6}
\hn\ce(\br)&=&Q\sum_{i=1}^{N\w}\left[\delta^3\left(\br-\bu_i-\frac{\ba_i}{2}\right)-\delta^3\left(\br-\bu_i+\frac{\ba_i}{2}\right)\right]\nonumber\\
&&+\sum_{j=1}^p\sum_{k=1}^{N_j}q_j\delta^3(\br-\bv_{jk}),\\
\label{eq7}
\hn\h(\br)&=&\sum_{i=1}^{N\w}\delta^3(\br-\bu_i)+\sum_{j=1}^p\sum_{k=1}^{N_j}\delta^3(\br-\bv_{jk}).
\eea
In addition, the functions $w\w(\bu_i,\ba_i)$ and $w_j(\bv_{jk})$ in Eq.~(\ref{eq5}) are the onsite potentials acting on the solvent molecules and the ions, respectively. These one-body potentials will enable us to relate the average density of the particles to their bulk fugacity and concentration. Finally, in Eq.~(\ref{eq1}), the self energy of the particles subtracted from the total energy $E$ is $\epsilon=N\w\epsilon\w+\sum_jN_j\epsilon_j$, with the self-energy components of the solvent molecules and salt ions given by $\beta\epsilon\w=v\h(0)/2+Q^2\left[v\ce(0)-v\ce(a)\right]$ and $\beta\epsilon_j=\left[v\h(0)+q_j^2v\ce(0)\right]/2$, respectively.

At this point, in Eq.~(\ref{eq1}), we introduce an Hubbard-Stratonovich transformation for each type of pairwise interaction,
\bea\label{eq10}
&&e^{-\frac{1}{2}\int\mathrm{d}^3\br\mathrm{d}^3\br'\hn_\alpha(\br)v_\alpha(\br-\br')\hn_\alpha(\br')}\\
&&=\int\mathcal{D}\phi_\alpha\;e^{-\frac{1}{2}\int\mathrm{d}^3\br\mathrm{d}^3\br'\phi_\alpha(\br)v_\alpha(\br-\br')\phi_\alpha(\br')},\nonumber
\eea
with the fluctuating potential $\phi_\alpha(\br)$ associated with the Coulomb ($\alpha={\rm c}$) or HC interactions ($\alpha={\rm h}$). Then, passing from the canonical to the grand-canonical ensemble via the transformation
\be
\label{eq11}
Z\g=\sum_{N\w=1}^\infty\frac{\Lambda\w^{N\w}}{N\w!}\prod_{j=1}^p\sum_{N_j=1}^\infty\frac{\Lambda_j^{N_j}}{N_j!}Z\ce,
\ee
the grand-canonical partition function takes the functional integral form 
\be
\label{eq12}
Z\g=\int\mathcal{D}\ph\;e^{-\beta H_{\rm el}[\ph]},
\ee
where we introduced the shortcut vector notations $\ph=(\phi\ce,\phi\h)$ and $\mathcal{D}\ph=\mathcal{D}\phi\ce\mathcal{D}\phi\h$. 

In Eq.~(\ref{eq12}), the electrolyte Hamiltonian
\be
\label{eq13}
H_{\rm el}[\ph]=H\w[\ph]+H\s[\ph]
\ee
is composed of the quadratic potential fluctuations and the contribution from the solvent molecules,
\bea
\label{eq14}
\beta H\w[\ph]&=&\frac{1}{2}\sum_{\alpha={\rm c,h}}\int\mathrm{d}^3\br\mathrm{d}^3\br'\phi_\alpha(\br)v_\alpha(\br-\br')\phi_\alpha(\br')\nonumber\\
&&-\int\frac{\mathrm{d}^2\bo}{4\pi}\mathrm{d}^3\bu\;\hd\w(\bu,\ba),
\eea
and the component associated with the salt ions,
\be
\label{eq15}
\beta H\s[\ph]=-\sum_{j=1}^p\int\mathrm{d}^3\bv\;\hd_j(\bv).
\ee
In Eqs.~(\ref{eq14}) and~(\ref{eq15}), we introduced the number density of the solvent and salt particles as
\bea
\label{eq16}
\hd\w(\bu,\ba)&=&\Lambda\w e^{-w\w(\bu,\ba)+i\phi\h(\bu)}e^{iQ\left[\phi\ce(\bu+\ba/2)-\phi\ce(\bu-\ba/2)\right]},\nonumber\\
&&\\
\label{eq16II}
\hd_j(\bv)&=&\Lambda_j e^{-w_j(\bv)+i\phi\h(\bv)}e^{iq_j\phi\ce(\bv)}.
\eea

\subsection{Introducing the perturbative schemes}
\label{pr}

Due to the non-linear functional form of  the solvent and salt densities in Eqs.~(\ref{eq14})-(\ref{eq15}), the partition function~(\ref{eq12}) cannot be evaluated analytically. In this section, we introduce a mixed perturbative treatment of the solvent and salt particles. This perturbative scheme will enable us the analytical evaluation of the salt-dressed solvent permittivity from the partition function~(\ref{eq12}).

\subsubsection{Virial approximation for salt ions}

In biological solutions, the typical salt concentrations located in the submolar range are by two orders of magnitude lower than the water concentration. This implies that the salt Hamiltonian~(\ref{eq15}) is the dilute component of the total Hamiltonian~(\ref{eq13}). Based on this point, we will treat the salt contribution~(\ref{eq15}) within a virial approximation. Thus, the remainder of the work will be limited to the linear order in the salt fugacity $O(\Lambda_j)$ or equivalently the salt concentration $O(\rho_j)$. The corresponding virial approximation will enable us to avoid the gaussian-level treatment of the ion density that would limit the theory to the weak electrostatic coupling regime where the dielectric decrement by pure electrostatic ion-solvent interactions was observed to be absent~\cite{Andelman2013,Andelman2018,JCP2013}.

In the grand-canonical ensemble, the statistical average of the general functional $F[\ph]$ is defined as
\be\label{eq17}
\lan F[\ph]\ran=\frac{1}{Z_{\rm G}}\int\mathcal{D}\ph\;e^{-\beta H_{\rm el}[\ph]}F[\ph].
\ee
Following the dilute salt approximation, we Taylor-expand Eqs.~(\ref{eq12}) and~(\ref{eq17}) at the linear order in the salt Hamiltonian~(\ref{eq15}). Upon this virial expansion, the field average~(\ref{eq17}) becomes
\bea\label{eq18}
\lan F[\ph]\ran&\approx&\lan F[\ph]\ran\w\\
&&-\left\{\lan\beta H\s[\ph]F[\ph]\ce\ran\w-\lan\beta H\s[\ph]\ran\w\lan F[\ph]\ce\ran\w\right\},\nonumber
\eea
where we defined the statistical average over the solvent degrees of freedom governed by the Hamiltonian~(\ref{eq14}),
\be\label{eq19}
\lan F[\ph]\ran\w=\frac{1}{Z\w}\int\mathcal{D}\ph\;e^{-\beta H\w[\ph]}F[\ph],
\ee
with the solvent partition function $Z\w=\int\mathcal{D}\ph\;e^{-\beta H\w[\ph]}$.

\subsubsection{Gaussian approximation for solvent particles}

The non-linear dependence of the solvent Hamiltonian~(\ref{eq14}) on the fluctuating potentials $\phi\ce(\br)$ and $\phi\h(\br)$ does not allow the analytical evaluation of the field-theoretic averages in Eq.~(\ref{eq19}) over the solvent fluctuations. To progress further, we will approximate the solvent Hamiltonian by the following functional of quadratic dependence on the potentials $\phi\ce(\br)$ and $\phi\h(\br)$,
\bea
\label{eq20}
&&H_w\approx\int\frac{\mathrm{d}^3\br\mathrm{d}^3\br'}{2}\left\{(\phi\ce-i\phi_0)_\br v^{-1}_0(\br-\br')(\phi\ce-i\phi_0)_{\br'}\right.\nonumber\\
&&\hspace{2.9cm}\left.+\phi\h(\br)v^{-1}\h(\br-\br')\phi\h(\br')\right\}.
\eea
In Eq.~(\ref{eq20}), we neglected the direct coupling between the fluctuating HC and electrostatic potentials, as well as the renormalization of the bare HC interaction potential $v\h(\br)$ by many-body particle collisions. Moreover, we introduced the electrostatic potential $\phi_0(\br)$ and the solvent kernel $v_0(\br-\br')$  that will be self-consistently determined from the solution of the SD equations. 

Due to the quadratic form of the functional Hamiltonian~(\ref{eq20}), the calculation of the statistical average in Eq.~(\ref{eq18}) reduces to the evaluation of Gaussian functional integrals. To this aim, we will extensively use the identity
\bea
\label{eq20II}
&&\lan e^{i\int\mathrm{d}\br\left[J\ce(\br)\phi\ce(\br)+J\h(\br)\phi\h(\br)\right]}\ran\w\\
&&=e^{-\int\mathrm{d}\br J\ce(\br)\phi_0(\br)}\prod_{\alpha={\rm c,h}}e^{-\frac{1}{2}\int_{\br,\br'}J_\alpha(\br)v_\alpha(\br-\br')J_\alpha(\br')}\nonumber
\eea
including the generating functions $J_{\rm c,h}(\br)$.

The remainder of the present work will be based on the solution of the electrostatic SD equations
\bea\label{SD1}
&&\lan\frac{\delta\left(\beta H_{\rm el}[\ph]\right)}{\delta\phi\ce(\br)}\ran=0,\\
\label{SD2}
&&\lan\frac{\delta\left(\beta H_{\rm el}[\ph]\right)}{\delta\phi\ce(\br)}\phi\ce(\br')\ran=\delta(\br-\br'),
\eea
where the field averages will be calculated within the approximation schemes introduced above. Eq.~(\ref{SD1}) is a beyond-mean-field (MF) Poisson identity, and Eq.~(\ref{SD2}) corresponds to a kernel equation characterizing the electrostatic fluctuations in the liquid. The derivation of these formally exact identities is reported in Appendix~\ref{ap1}.

\subsection{Calculation of the particle densities and the electroneutrality condition}

In this part, we calculate first the average density of the ions and the solvent molecules, and relate their fugacity to their bulk concentration. Then, by evaluating the SD Eq.~(\ref{SD1}), we show that the mixed perturbation scheme introduced in Ref.~\ref{pr} consistently satisfies the global electroneutrality condition. With the aim to simplify the notation, from now on, the argument of the Hamiltonian functionals will be omitted.

\subsubsection{Relating the particle fugacities and concentrations} 

The average density of the salt ions follows from the partition function~(\ref{eq12}) via the identity $\rho_j=-Z_{\rm G}^{-1}\delta Z_{\rm G}/\delta w_j(\br)=\lan \hd_j(\bv)\ran$. Substituting Eq.~(\ref{eq16II}) into Eq.~(\ref{eq18}), using the identity~(\ref{eq20II}), and setting the onsite potential $w_j(\br)$ to zero, the ion density follows at the order $O(\Lambda_j)$ as $\rho_j=\Lambda_je^{-q_j^2\left[v_0(0)-v_c(0)\right]-q_j\phi_0(\br)}$. Moreover, noting that the uniformity of the bulk ion concentration implies a uniform potential $\phi_0(\br)=\phi_0$, the relation between the ion concentration and fugacity follows as
\be\label{eq22}
\Lambda_j=\rho_je^{q_j^2\left[v_0(0)-v_c(0)\right]+q_j\phi_0}.
\ee

In order to calculate the average solvent density, we set $w\w(\br,\ba)=w\w(\br)$ and use the identity $\rho\w=-Z^{-1}_{\rm G}\delta Z_{\rm G}/\delta w\w(\br)$ to get
\be
\label{eq23}
\rho\w=\int\frac{\mathrm{d}^2\bom}{4\pi}\lan \hd\w(\bu,\ba)\ran.
\ee
Plugging the function~(\ref{eq16}) into Eq.~(\ref{eq18}), and evaluating the field-theoretic averages with the identity~(\ref{eq20II}), after lengthy algebra, Eq.~(\ref{eq23}) yields at the order $O(\rho_j)$
\be
\label{eq24}
\Lambda\w=\rho\w e^{Q^2\left[v_0(0)-v\ce(0)-v_0(a)+v\ce(a)\right]}\left(1-\delta\rho\right).
\ee
In Eq.~(\ref{eq24}), the second virial coefficient originating from the salt-solvent interactions reads
\be
\label{eq25}
\delta\rho=\sum_{j=1}^p\rho_j\int\mathrm{d}^3\bk\;h_j(\bk,\ba),
\ee
where we introduced the Mayer function
\be
\label{eq26}
h_j(\bk,\ba)=e^{-v\h(\bk)-q_jQ\left[v_0(\bk-\ba/2)-v_0(\bk+\ba/2)\right]}-1,
\ee
and the C.M. distance $\bk=\bv-\bu$ between the salt ions and the solvent molecules. Finally, in order to evaluate the density of the terminal charges of the solvent molecules, we redefine the dipolar onsite potential as $w\w(\bu,\ba)=w_{{\rm w}+}(\bu+\ba/2)+w_{{\rm w}-}(\bu-\ba/2)$. Using the identities $\rho_{{\rm w}\pm}=-Z_{\rm G}^{-1}\delta Z_{\rm G}/\delta w_{{\rm w}\pm}(\br)$, after some algebra, one obtains the expected result 
\be\label{eq27}
\rho_{{\rm w}\pm}=\rho\w.
\ee

\subsubsection{Electroneutrality condition from the Poisson Eq.~(\ref{SD1})}

In this part, we show that the global electroneutrality condition consistently follows from the beyond-MF Poisson Eq.~(\ref{SD1}). First, by combining Eqs.~(\ref{eq18}) and~(\ref{SD1}), the Poisson Eq. becomes at the order $O(\rho_j)$
\bea\label{eq29}
&&\lan\frac{\delta\left(\beta H\w\right)}{\delta\phi\ce(\br)}\ran\w+\lan\frac{\delta\left(\beta H\s\right)}{\delta\phi\ce(\br)}\ran\w\\
&&-\lan\beta H\s\frac{\delta\left(\beta H\w\right)}{\delta\phi\ce(\br)}\ran\w+\lan\beta H\s\ran\w\lan\frac{\delta\left(\beta H\w\right)}{\delta\phi\ce(\br)}\ran\w\nonumber=0.
\eea
Next, we substitute into Eq.~(\ref{eq29}) the Hamiltonian components~(\ref{eq14})-(\ref{eq15}), and make use of the identity~(\ref{eq20II}) to evaluate the functional averages over the solvent fluctuations. Using as well Eqs.~(\ref{eq22}), (\ref{eq24}), and~(\ref{eq27}), after long algebra, Eq.~(\ref{eq29}) becomes
\be
\label{eq30}
\int\mathrm{d}^3\br'v\ce^{-1}(\br-\br')\left[\phi_0(\br')+\phi\s(\br')\right]-\sum_{j=1}^pq_j\rho_j=0,
\ee
where we defined the average potential component induced by the salt ions, 
\be
\label{eq31}
\phi\s(\br)=\sum_{j=1}^p\rho_jq_j\int\mathrm{d}^3vv_0(\br-\bv).
\ee
Injecting into Eqs.~(\ref{eq30})-(\ref{eq31}) the Fourier transform of the Green's functions
\be\label{eq32}
v_\alpha(\br-\br')=\int\frac{\mathrm{d}^3\bq}{(2\pi)^3}\;e^{i\bq\cdot\left(\br-\br'\right)}\tv_\alpha(q)
\ee
for $\alpha=\{0,{\rm c}\}$, with the Fourier-transformed bulk Coulomb potential $\tv\ce(q)=q^2/(4\pi\ell_{\rm B})$ in vacuum, and accounting for the uniformity of the average potentials, i.e. $\phi_0(\br)=\phi_0$ and $\phi\s(\br)=\phi\s$, Eq.~(\ref{eq30}) finally reduces to the electroneutrality condition for the salt ions, i.e.
\be\label{eq33}
\sum_{j=1}^p\rho_jq_j=0.
\ee

\subsection{Calculation of the salt-dressed solvent permittivity from the electrostatic kernel}

Here, we calculate the electrostatic Green's function and obtain the dielectric permittivity of the solvent. 

\subsubsection{Derivation of the electrostatic kernel equation}

The electrostatic Green's function is defined as the variance of the electrostatic potential fluctuations,
\be
\label{eq34}
G(\br-\br')=\lan\phi(\br)\phi(\br')\ran-\lan\phi(\br)\ran\lan\phi(\br')\ran.
\ee
Using the identity~(\ref{eq18}) for the virial-expanded field average, at the order $O(\rho_j)$, Eq.~(\ref{eq34}) takes the form
\bea\label{eq35}
G(\br-\br')&=&\lan\phi(\br)\phi(\br')\ran\w-\lan\phi(\br)\ran\w\lan\phi(\br')\ran\w\\
&&+\lan\phi(\br)\phi(\br')\ran\w\lan\beta H\s\ran\w-\lan\phi(\br)\phi(\br')\beta H\s\ran\w\nonumber\\
&&+\left[\lan\phi(\br)\beta H\s\ran\w-\lan\phi(\br)\ran\w\lan\beta H\s\ran\w\right]\lan\phi(\br')\ran\w\nonumber\\
&&+\left[\lan\phi(\br')\beta H\s\ran\w-\lan\phi(\br')\ran\w\lan\beta H\s\ran\w\right]\lan\phi(\br)\ran\w.\nonumber
\eea
Substituting now the salt Hamiltonian~(\ref{eq15}) into Eq.~(\ref{eq35}), and evaluating with Eq.~(\ref{eq20II}) the averages over the solvent configurations, the net Green's function becomes
\be
\label{eq36}
G(\br-\br')=v_0(\br-\br')-\sum_{j=1}^p\rho_jq_j^2\int\mathrm{d}^3\bv\;v_0(\br-\bv)v_0(\bv-\br').
\ee
Finally, by Fourier-transforming Eq.~(\ref{eq36}) according to Eq.~(\ref{eq32}), the inverse kernel follows at the order $O(\rho_j)$ as
\be
\label{eq37}
\tG^{-1}(q)=\tv_0^{-1}(q)+\sum_{j=1}^p\rho_jq_j^2.
\ee

The first term on the r.h.s. of Eq.~(\ref{eq37}) embodies the dielectric screening induced solely by the solvent molecules, and its modification by the salt-solvent scattering. Then, the second term accounts for the direct Debye screening by the salt charges. In order to characterize the effect of salt on the solvent permittivity, we calculate now the kernel equation~(\ref{SD2}) satisfied by the potential  $v_0(\br-\br')$.

Plugging Eq.~(\ref{eq13}) into the l.h.s. of Eq.~(\ref{SD2}), and using Eq.~(\ref{eq18}), at the order $O(\rho_j)$, the kernel equation becomes
\bea\label{eq38}
&&\lan\frac{\delta\left(\beta H\w\right)}{\delta\phi\ce(\br)}\phi\ce(\br')\ran\w+\lan\frac{\delta\left(\beta H\s\right)}{\delta\phi\ce(\br)}\phi\ce(\br')\ran\w\\
&&-\lan\beta H\s\frac{\delta\left(\beta H\w\right)}{\delta\phi\ce(\br)}\phi\ce(\br')\ran\w+\lan\beta H\s\ran\w\lan\frac{\delta\left(\beta H\w\right)}{\delta\phi\ce(\br)}\phi\ce(\br')\ran\w\nonumber\\
&&=\delta(\br-\br').\nonumber
\eea
Next, we plug into Eq.~(\ref{eq38}) the Hamiltonian components~(\ref{eq14})-(\ref{eq15}), and evaluate the functional averages over the solvent degrees of freedom with the identity~(\ref{eq20II}). After long but straigthforward algebra, the Green equation satisfied by the solvent kernel $v_0(\br-\br')$ finally follows as
\begin{widetext}
\bea
\label{eq39}
&&\int\mathrm{d}^3\br_1v\ce^{-1}(\br-\br_1)v_0(\br_1-\br')+\sum_{j=1}^p\rho_jq_j^2v_0(\br-\br')
+Q^2\rho\w\int\frac{\mathrm{d}^2\bo}{4\pi}\left[2v_0(\br-\br')-v_0(\br-\ba-\br')-v_0(\br+\ba-\br')\right]\nonumber\\
&&+Q\rho\w\sum_{j=1}^p\rho_jq_j\int\frac{\mathrm{d}^2\bo}{4\pi}\mathrm{d}^3\bv\left[h_j\left(\bv-\br+\frac{\ba}{2},\ba\right)-h_j\left(\bv-\br-\frac{\ba}{2},\ba\right)\right]v_0(\bv-\br')\nonumber\\
&&-\sum_{j=1}^p\rho_jq_j^2\int\mathrm{d}^3\bv\mathrm{d}^3\br_1v\ce^{-1}(\br-\br_1)v_0(\br_1-\bv)v_0(\bv-\br')=\delta(\br-\br').
\eea
\end{widetext}

For the sake of consistency with the perturbative treatment of salt, Eq.~(\ref{eq39}) should be linearized in the salt concentration. To this aim, we formally express the solution to Eq.~(\ref{eq39}) as
\be
\label{eq40}
v_0(\br-\br')=v\w(\br-\br')+\lambda v\s(\br-\br'),
\ee
where the kernel component $v\w(\br-\br')$ is associated with pure solvent, and the correction term $v\s(\br-\br')$ accounts for the solvent-salt correlations. In Eq.~(\ref{eq40}), we introduced the dimensionless perturbative parameter $\lambda=O(\rho_j)$ that will be set to unity at the end of the calculation. Plugging Eq.~(\ref{eq40}) into Eq.~(\ref{eq39}), and expanding the result at the order $O(\rho_j)$, one obtains the following integral equations satisfied by each kernel component,
\begin{widetext}
\bea
\label{eq41}
&&\int\mathrm{d}^3\br_1v\ce^{-1}(\br-\br_1)v\w(\br_1-\br')+Q^2\rho\w\int\frac{\mathrm{d}^2\bo}{4\pi}\left[2v\w(\br-\br')-v\w(\br-\ba-\br')-v\w(\br+\ba-\br')\right]=\delta(\br-\br'),\\
\label{eq42}
&&\int\mathrm{d}^3\br_1v\ce^{-1}(\br-\br_1)v\s(\br_1-\br')+Q^2\rho\w\int\frac{\mathrm{d}^2\bo}{4\pi}\left[2v\s(\br-\br')-v\s(\br-\ba-\br')-v\s(\br+\ba-\br')\right]\nonumber\\
&&+Q\rho\w\sum_{j=1}^p\rho_jq_j\int\frac{\mathrm{d}^2\bo}{4\pi}\mathrm{d}^3\bv\left[h_j\left(\bv-\br+\frac{\ba}{2},\ba\right)-h_j\left(\bv-\br-\frac{\ba}{2},\ba\right)\right]v\w(\bv-\br')
+\sum_{j=1}^p\rho_jq_j^2v\w(\br-\br')\nonumber\\
&&-\sum_{j=1}^p\rho_jq_j^2\int\mathrm{d}^3\bv\mathrm{d}^3\br_1v\ce^{-1}(\br-\br_1)v\w(\br_1-\bv)v\w(\bv-\br')=0,
\eea
\end{widetext}
with the expanded Mayer function 
\be\label{m2}
h_j(\bk,\ba)=e^{-v\h(\bk)-q_jQ\left[v\w(\bk-\ba/2)-v\w(\bk+\ba/2)\right]}-1. 
\ee

\subsubsection{Solution of the kernel equations~(\ref{eq41})-(\ref{eq42})}

Eq.~(\ref{eq41}) is an integro-differential equation satisfied by the salt-free solvent kernel~\cite{JCP2013}. By Fourier-transforming this identity according to Eq.~(\ref{eq32}), and evaluating the integral over the solid angle $\bo$, the solution directly follows as the non-local Coulomb law firstly derived in Ref.~\cite{JCP2013},
\bea
\label{eq43}
\tv\w^{-1}(q)=\frac{q^2\te_{\rm w}(q)}{4\pi\ell_{\rm B}};\hspace{2mm}\te_{\rm w}(q)=1+\frac{\kappa\w^2}{q^2}\left[1-\frac{\sin(qa)}{qa}\right].
\eea
In Eq.~(\ref{eq43}), we introduced the dielectric permittivity spectrum $\te_{\rm w}(q)$ with the dielectric screening parameter 
\be\label{kapw}
\kappa\w=\sqrt{8\pi Q^2\ell_{\rm B}\rho\w}.
\ee
In the infrared (IR) limit, this permittivity function tends to the static dielectric permittivity of the pure solvent, i.e. $\te\w(q\to0)=\e_{\rm w,b}$, with the bulk solvent permittivity given by the  Langevin equation
\be
\label{lang}
\e_{\rm w,b}=1+\frac{4\pi}{3}\ell_{\rm B}Q^2a^2\rho\w.
\ee
Unless stated otherwise, in the remainder, we will set the dipole length to the value $a=1$ {\AA} yielding the water permittivity $\e_{\rm w,b}\approx78.8$ at the temperature $T=298$ K.

In order to derive the correction kernel $v\s(\br-\br')$, we plug into Eq.~(\ref{eq42}) the Fourier expansion of the Green's functions in Eq.~(\ref{eq32}). Carrying out the spatial and angular integrals, after some algebra, the Fourier-transformed correction kernel follows as
\be
\label{eq44}
\tv\s(q)=-\sum_{j=1}^p\rho_jq_j\left\{q_j\left[1-\te^{-1}_{\rm w}(q)\right]+Q\rho\w T_j(q)\right\}\tv\w^2(q),\nonumber
\ee
with the auxiliary integral
\be
\label{eq45}
T_j(q)=\int\frac{\mathrm{d}^2\bo}{4\pi}\mathrm{d}^3\bk\left[h_j(\bk+\frac{\ba}{2},\ba)-h_j(\bk-\frac{\ba}{2},\ba)\right]e^{i\bq\cdot\bk}.
\ee
Hence, the Fourier-transformed inverse of Eq.~(\ref{eq40}) given by $\tv_0^{-1}(q)=\tv\w^{-1}(q)-\tv\w^{-2}(q)\tv\s(q)+O(\rho_j^2)$ becomes
\be
\label{eq46}
\tv_0^{-1}(q)=\frac{q^2\te_{\rm w}(q)}{4\pi\ell_{\rm B}}+\sum_{j=1}^p\rho_jq_j\left\{q_j\left[1-\te^{-1}_{\rm w}(q)\right]+Q\rho\w T_j(q)\right\}.
\ee

\subsubsection{Computation of the salt-dressed dielectric permittivity}

Injecting now Eq.~(\ref{eq46}) into Eq.~(\ref{eq37}), the inverse total kernel follows as
\be\label{eq46II}
\tG^{-1}(q)=\frac{1}{4\pi\ell\w}\sum_{j=1}^p\left[\kappa_j^2+\delta\kappa_j^2\right]+\frac{q^2\te(q)}{4\pi\ell_{\rm B}},
\ee
with the Fourier transform of the electrolyte permittivity
\be\label{eq46III}
\te(q)=\te_{\rm w}(q)+\delta\te(q).
\ee
In Eq.~(\ref{eq46II}), we introduced the DH parameter associated with each ion species as $\kappa_j^2=4\pi\ell\w\rho_jq_j^2$, with the Bjerrum length in water $\ell\w=\ell_{\rm B}/\e_{\rm w,b}$. Moreover, we defined the following correction terms embodying the effect of the solvent-salt correlations on the Debye screening and the dielectric permittivity,
\bea
\label{eq47}
\hspace{-3mm}&&\delta\kappa_j^2=\kappa_j^2\left[\frac{Q}{\rho\w}{q_j}T_j(0)+\frac{\e_{\rm w,b}-1}{\e_{\rm w,b}}\right],\\
\label{eq48}
\hspace{-3mm}&&\delta\te(q)=\sum_{j=1}^p\frac{\kappa_j^2}{q^2}\left\{\frac{\e_{\rm w,b} Q\rho\w}{q_j}\left[T_j(q)-T_j(0)\right]\right.\nonumber\\
&&\left.\hspace{2.5cm}+1-\frac{\e_{\rm w,b}}{\te_{\rm w}(q)}\right\}.
\hspace{-3mm}
\eea

\begin{figure*}
\includegraphics[width=1.\linewidth]{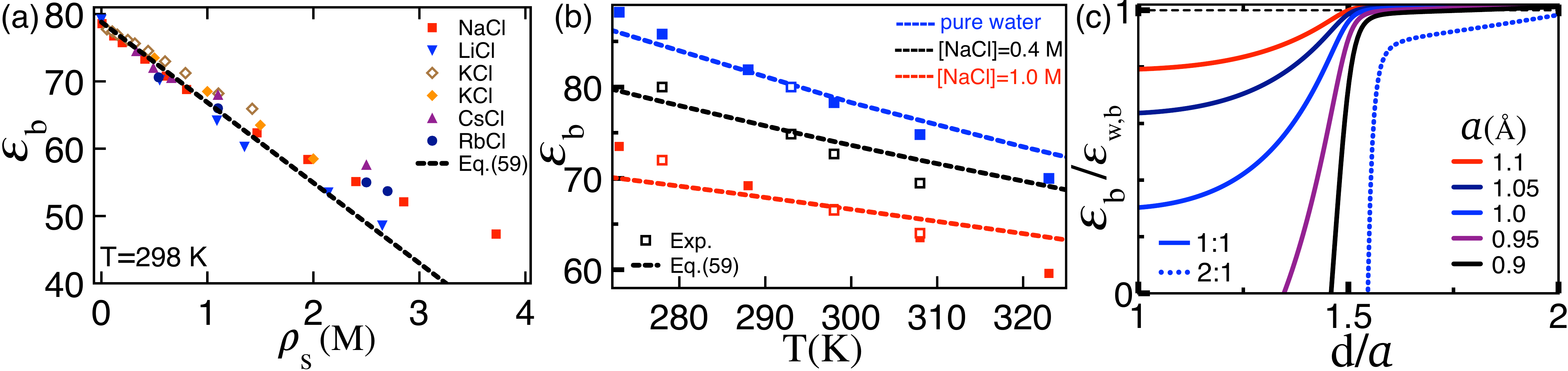}
\caption{(Color online) (a)  Dielectric permittivity of 1:1 solutions versus the salt concentration at $T=298$ K. (b) NaCl and water permittivities versus temperature. The HC diameter in (a)-(b) is $d=1.485$ {\AA}. (c) Permittivity of the 1:1 (solid curves) and 2:1 solutions (dotted curve) versus the HC diameter at the salt concentration $\rho_j=0.5$ M, temperature $T=298$ K, and various solvent diameters. In all plots, the curves display Eq.~(\ref{eq55}), and the symbols are the experimental data. In (a), the data for NaCl are from Ref.~\cite{Buchner1999},  KCl from Ref.~\cite{Hasted1948} (filled diamond symbols) and~\cite{Chen2003} (open diamond symbols), LiCl from Ref.~\cite{Sridhar1990}, and CsCl and RbCl from Ref.~\cite{Wei1992}. The data in (b) are from Ref.~\cite{Buchner1999} (open squares) and Ref.~\cite{Pottel} (filled squares).}
\label{fig2} 
\end{figure*}

For the evaluation of Eqs.~(\ref{eq47})-(\ref{eq48}), we reduce the integral~(\ref{eq45}) to a numerically manageable form. Substituting the Mayer function~(\ref{m2}) into Eq.~(\ref{eq45}), the latter becomes
\be
\label{eq49}
T_j(q)=\int\mathrm{d}^3\bk e^{i\bq\cdot\bk}K_j(q,\bk),
\ee
where we introduced the auxiliary integral
\bea
\label{eq49II}
K_j(q,\bk)&=&\int\frac{\mathrm{d}^2\bo}{4\pi}\left\{e^{-v\h(\bk+\ba/2)-q_jQ\left[v\w(\bk)-v\w(\bk+\ba)\right]}\right.\\
&&\hspace{1.3cm}\left.-e^{-v\h(\bk-\ba/2)+q_jQ\left[v\w(\bk)-v\w(\bk-\ba)\right]}\right\}.\nonumber
\eea
Noting that the HC and Coulomb potentials in Eq.~(\ref{eq49II}) depend exclusively on the magnitude of their argument, this integral can be expressed as
\bea
\label{eq50}
K_j(q,\bk)&=&\int\frac{\mathrm{d}^2\bo}{4\pi}\left\{e^{-v\h(u_+)-q_jQ\left[v\w(k)-v\w(t_+)\right]}\right.\\
&&\hspace{1.7cm}\left.-e^{-v\h(u_-)+q_jQ\left[v\w(k)-v\w(t_-)\right]}\right\},\nonumber
\eea
where we defined the variables $u_\pm=\sqrt{k^2+a^2/4\pm\bk\cdot\ba}$ and $t_\pm=\sqrt{k^2+a^2\pm2\bk\cdot\ba}$. Considering now the specific HC interaction potential defined by Eq.~(\ref{hcp}), and expressing the integral~(\ref{eq50}) in terms of the spherical coordinates associated with a Cartesian coordinate system whose $z$ axis aligns with the vector $\bk$, one obtains
\bea
\label{eq51}
K_j(q,k)&=&\int_0^\pi\frac{\mathrm{d}\theta}{2}\sin\theta\left\{e^{-q_jQ\left[v\w(k)-v\w(t_+)\right]}H(u_+-d)\right.\nonumber\\
&&\hspace{1.9cm}\left.-e^{q_jQ\left[v\w(k)-v\w(t_-)\right]}H(u_--d)\right\},\nonumber\\
\eea
where $u_\pm=\sqrt{k^2+a^2/4\pm ka\cos\theta}$ and $t_\pm=\sqrt{k^2+a^2\pm2ka\cos\theta}$. Finally, changing the integration variable in Eq.~(\ref{eq51}) from $\theta$ to $t_\pm$ for the first term ($\theta\to t_+$) and the second term ($\theta\to t_-$), and plugging the resulting integral into Eq.~(\ref{eq49}), one finally obtains
\bea\label{eq52}
T_j(q)&=&\frac{4\pi}{a}\int_{d-\frac{a}{2}}^\infty\mathrm{d}kk\frac{\sin(qk)}{qk}\\
&&\hspace{0mm}\times\int_{t_<(k)}^{k+a}\mathrm{d}tt\sinh\left\{Qq_j\left[v\w(t)-v\w(k)\right]\right\},\nonumber
\eea
with the cut-off distance accounting for the HC interactions $t_<(k)={\rm max}\left(k-a,\sqrt{2d^2-k^2+a^2/2}\right)$.

The static permittivity of the electrolyte corresponds to the IR limit of the Fourier-transformed permittivity function~(\ref{eq46III}), i.e.  $\e_{\rm b}=\te(q\to0)=\e_{\rm w,b}+\delta\te(q\to0)$. Substituting the integral~(\ref{eq52}) into Eqs.~(\ref{eq47})-(\ref{eq48}), and taking the IR limit, the correction to the screening parameter and the electrolyte permittivity finally follow as
\bea
\label{eq54}
\delta\kappa_j^2&=&\kappa_j^2\left\{\frac{\e_{\rm w,b}-1}{\e_{\rm w,b}}+\frac{4\pi Q\rho\w}{aq_j}I^{(1)}_j\right\},\\
\label{eq55}
\e_{\rm b}&=&\e_{\rm w,b}-\frac{\pi}{3Q}\frac{\kappa\w^2}{a}\sum_{j=1}^p\rho_jq_jI^{(3)}_j,
\eea
where we introduced the auxiliary integrals
\be\label{eq56}
I^{(n)}_j=\int_{d-\frac{a}{2}}^\infty\mathrm{d}kk^n\int_{t_<(k)}^{k+a}\mathrm{d}tt\sinh\left\{Qq_j\left[v\w(t)-v\w(k)\right]\right\}
\ee
embodying the HC and electrostatic interactions between the salt charges and the solvent molecules. 

At this point, we note that the numerical computation of the integrals in Eq.~(\ref{eq56}) requires the knowledge of the non-local electrostatic potential $v\w(r)$ in the pure solvent. This potential follows from Eqs.~(\ref{eq32}) and~(\ref{eq43}) as
\be
\label{nlvc}
v\w(r)=\frac{\ell_{\rm B}}{\e\w(r)r}\;;\hspace{5mm}\e\w(r)=\frac{\pi}{2}\left\{\int_0^\infty\frac{\mathrm{d}q}{q}\frac{\sin (qr)}{\te_{\rm w}(q)}\right\}^{-1},
\ee
where $\e\w(r)$ is the non-uniform permittivity characterizing the dielectric screening induced by the structured solvent at a distance $r$ from a charge source~\cite{JCP2013}. As this distance increases, the non-uniform permittivity tends to the bulk value given by the Langevin equation~(\ref{lang}), i.e. $\e\w(r\to\infty)=\e_{\rm w,b}$ (see Fig.~\ref{fig4}(c) below).

From the numerical evaluation of the integral $I_j^{(1)}$ in the second term of Eq.~(\ref{eq54}), we found that for salts of arbitrary composition, the correction to the Debye screening in Eq.~(\ref{eq46II}) is exactly zero~\cite{rem1}, i.e.
\be\label{eq57}
\sum_{j=1}^p\delta\kappa_j^2=0.
\ee
Thus, at the leading order, the ion-solvent correlations modify exclusively the solvent-induced dielectric screening without affecting the salt-driven Debye screening. 

We finally calculate the modification of the solvent permittivity profile in Eq.~(\ref{nlvc}) by added salt. From Eqs.~(\ref{eq37}) and~(\ref{eq46II}), the inverse Fourier-transformed electrostatic interaction potential excluding the Debye screening follows as $\tv_0^{-1}(q)=q^2\te(q)/(4\pi\ell_{\rm B})$. Passing from the reciprocal to the real space via an inverse Fourier transform, and expanding the result at the linear order in the salt concentration, the non-local electrostatic potential and the salt-dressed permittivity profile become 
\be\label{elpr}
v_0(r)=\frac{\ell_{\rm B}}{\e(r)r}\;;\hspace{5mm}\e(r)=\e\w(r)+\delta\e(r)+O\left(\rho_j^2\right),
\ee
where the permittivity correction associated with the salt-solvent correlations reads
\be
\label{delpr}
\delta\e(r)=\frac{2\e\w^2(r)}{\pi}\int_0^\infty\mathrm{d}q\frac{\sin (qr)}{q}\frac{\delta\te(q)}{\te\w^2(q)}.
\ee

\section{Results and Discussion}
\label{res}

\subsection{Salt-induced dielectric decrement}

The permittivity formula in Eq.~(\ref{eq55}) is the key result of this article. One notes that this identity displays explicitly the experimentally observed linear decrement of the liquid permittivity with the dilute salt concentration $\rho_j$. Fig.~\ref{fig2}(a) illustrates the theoretical prediction~(\ref{eq55}) (dashed curve) and the experimental permittivity data of various monovalent salt solutions against the salt concentration $\rho_\pm=\rho\s$ at the liquid temperature $T=298$ K. The HC diameter is set to the value $d=1.485$ {\AA} yielding the best agreement with the slope of the NaCl permittivity data. The plot shows that the linear decrement of the permittivity occurs in the density regime $\rho\s\lesssim2.0$ M. 

In order to shed light on the nature of the particle correlations driving the dielectric decrement, we note that the integrals~(\ref{eq49})-(\ref{eq49II}), or equivalently the integral~(\ref{eq56}) in Eq.~(\ref{eq55}) correspond to the trace of the Mayer functions~(\ref{m2}) associated with the interacting salt ($q_j$) and solvent charges ($\pm Q$) over the ion and dipole configurations. Thus, according to Eq.~(\ref{eq55}), the dielectric decrement is the manifestation of the salt-induced electrostatic screening of the polarization charges reducing the dielectric response of the solvent. Next, we investigate the role of the liquid temperature, the HC and solvent radii, and the ion valency on this decrement mechanism.

\subsection{Effect of temperature}

\begin{figure}
\includegraphics[width=1.\linewidth]{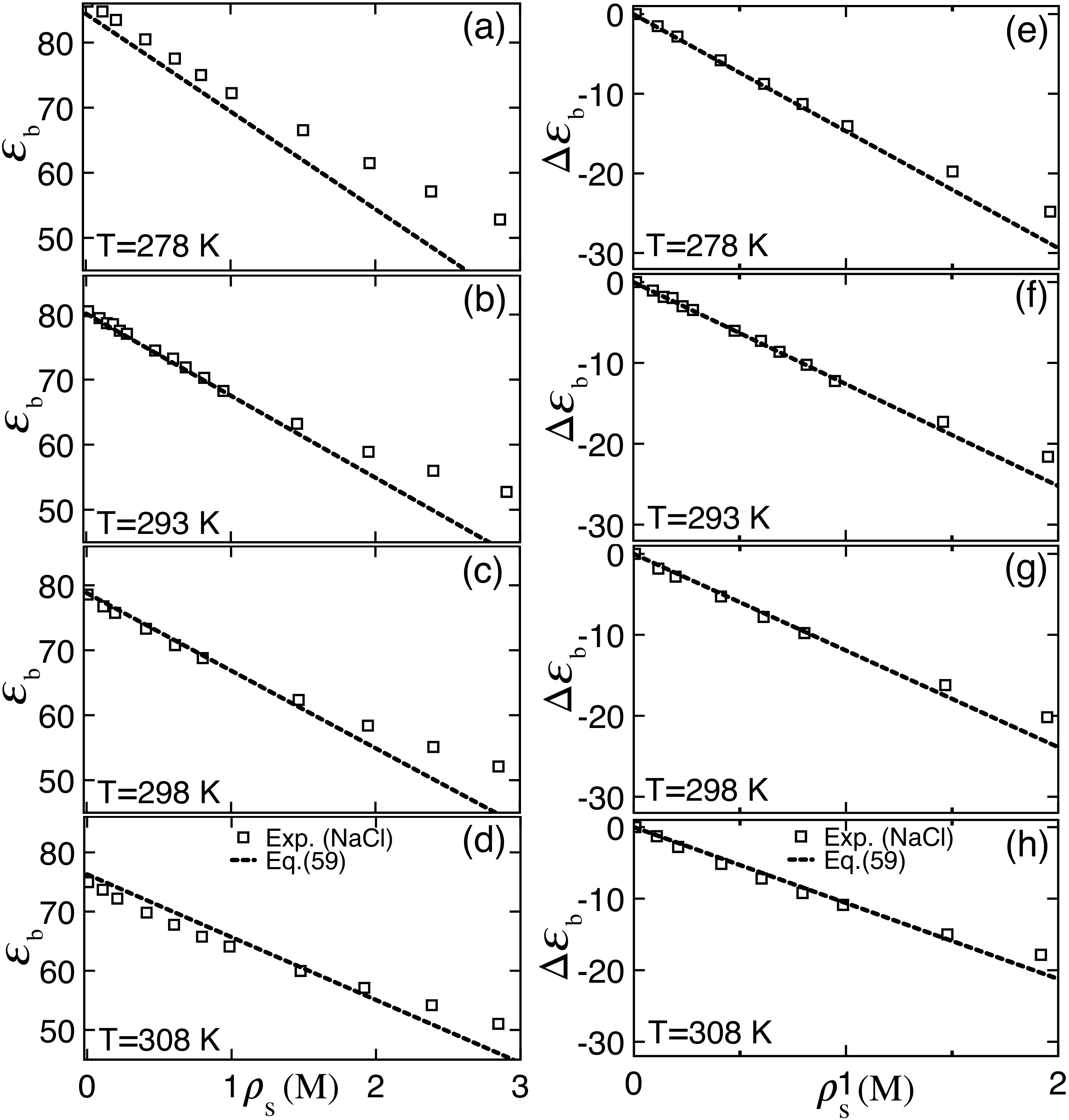}
\caption{(Color online) (a)-(d) Electrolyte permittivity and (e)-(h) reduced permittivity $\Delta\e_{\rm b}=\e_{\rm b}-\e_{\rm w,b}$ versus the NaCl concentration at various liquid temperatures. The HC size is $d=1.485$ {\AA}. The experimental data are from Ref.~\cite{Buchner1999}.}
\label{fig3} 
\end{figure}
\begin{figure*}
\includegraphics[width=1.\linewidth]{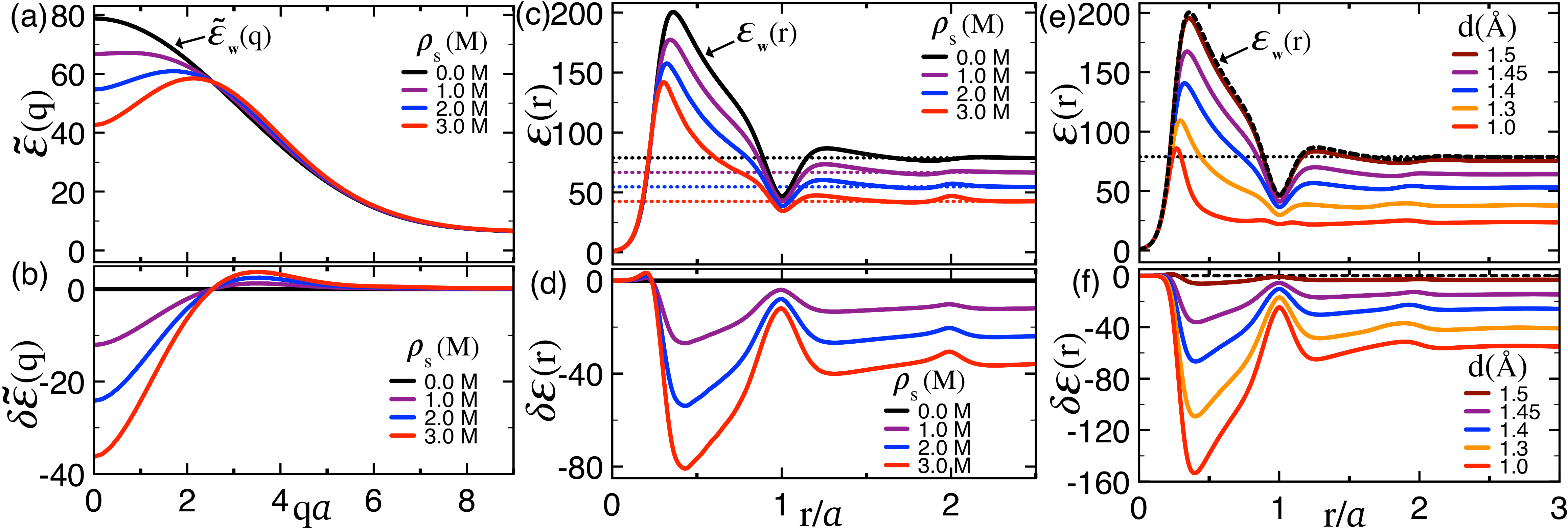}
\caption{(Color online) (a) Spectrum of the Fourier-transformed electrolyte permittivity~(\ref{eq46III}) and (b) its salt component~(\ref{eq48}), (c) the spatial profile of the electrolyte permittivity in Eq.~(\ref{elpr}) and (d) the salt contribution~(\ref{delpr}) at the HC diameter $d=1.485$ {\AA} and various salt concentrations indicated in the legend. (e)-(f) The plots in (c)-(d) at the salt concentration $\rho\s=0.5$ M and various HC sizes given in the legend. The temperature is $T=298$ K  in all plots.}
\label{fig4} 
\end{figure*}

The temperature dependence of the liquid permittivity is illustrated in Fig.~\ref{fig2}(b). One notes that at various salt concentrations, the permittivity formula~(\ref{eq55}) can qualitatively reproduce the decay of the experimental NaCl permittivity with the rise of the temperature, i.e. $T\uparrow\e_{\rm b}\downarrow$. This feature is equally illustrated in Figs.~\ref{fig3}(a)-(d) displaying the dielectric decrement by salt at various temperatures. One sees that at submolar salt concentrations, Eq.~(\ref{eq55}) can account for the permittivity decrease by the rise of the temperature with reasonable accuracy. 

In Fig.~\ref{fig2}(b), the qualitatively identical thermal decay of the water (blue) and electrolyte permittivities (black and red) indicates that this effect is mainly caused by the suppression of the dielectric response of the background solvent by the amplification of the thermal fluctuations. Figs.~\ref{fig2}(b) and~\ref{fig3}(a)-(d) also show that the thermal susceptibility of the experimental permittivity data is underestimated by Eq.~(\ref{eq55}). This point is elaborated further in the Conclusions.

In order to characterize the thermal susceptibility of the salt-induced dielectric decrement separately from the temperature dependence of the background water, in Figs.~\ref{fig3}(e)-(h), we reported the reduced permittivity $\Delta\e_{\rm b}\equiv\e_{\rm b}-\e_{\rm w,b}$ against the salt concentration at various temperatures. From the top to the bottom plot, the inspection of the experimental data and the theoretical curves with respect to the right axis shows that the rise of the temperature lowers the magnitude of the dielectric decrement, i.e. $T\uparrow|\Delta\e_{\rm b}| \downarrow$. According to Eq.~(\ref{eq55}), this feature stems from the weakening of the salt-solvent correlations by thermal fluctuations. Namely, in Eq.~(\ref{eq56}), the rise of the temperature reduces the magnitude of the dimensionless electrostatic potential $v\w(r)$ mediating the electrostatic salt-solvent interactions. This weakens the salt screening of the polarization charges and suppresses the resulting dielectric decrement. Finally, one notes that the thermal decay of the dielectric decrement is reproduced by Eq.~(\ref{eq55}) with a significantly better accuracy than the decay of the total permittivity $\e_{\rm w,b}$. This point is equally discussed in Conclusions.

\subsection{Effect of HC and solvent radii, and salt valency}

Fig.~\ref{fig2}(c) illustrates the influence of the HC radius, the solvent size, and the salt valency on the dielectric decrement. The plot is limited to the physical regime of positive dielectric permittivity below which the virial treatment of the salt ions at the basis of our formalism breaks down. The figure shows that the electrolyte permittivity exhibits a high sensitivity to the solvent and HC radii. 

First, one sees that at constant ratio $d/a$, the lower the solvent size, the stronger the dielectric decrement, i.e. $a\downarrow\e_{\rm b}/\e_{\rm w,b}\downarrow$. To understand this point, we note that according to Eq.~(\ref{lang}), the reduction of the solvent size $a$ or equivalently the dipole moment $p=Qa$ lowers the solvent permittivity $\e_{\rm w,b}$. This weakens the dielectric screening experienced by the interaction potential $v\w(r)$ in Eq.~(\ref{eq56}), amplifying the magnitude of the salt-solvent correlations and the resulting dielectric decrement. 

In Fig.~\ref{fig2}(c), the comparison of the solid and dotted blue curves shows that salt valency plays a similar role on dielectric decrement. Namely, in Eq.~(\ref{eq56}), a higher salt valency gives rise to stronger salt-solvent correlations. As a result, the addition of a 1:2 salt into water leads to a more efficient dielectric decrement than a monovalent 1:1 salt, i.e. $q_j\uparrow\e_{\rm b}\downarrow$. It is noteworthy that this effect has been observed in the experiments of Harris and O'Konski conducted with mono- and divalent salt solutions~\cite{Harris1957}.

Finally, Fig.~\ref{fig2}(c) shows that the electrolyte permittivity is a uniformly rising function of the HC size, i.e. $d\uparrow\e_{\rm b}\uparrow$. This feature stems from the overall attenuation of the salt-solvent correlations by the increase of the close contact distance. This suppresses the dielectric decrement and rises the permittivity of the 1:1 solutions towards the water permittivity $\e_{\rm w,b}$ reached at the contact distance $d\approx1.5\;a$.  In the case of the 1:2 solutions characterized by stronger salt-solvent coupling (dotted curve), the dielectric decrement exhibits a sharper decay with the rise of the HC diameter, and the water permittivity is reached at the larger HC size of $d\approx2\;a$.
 
 \subsection{Spatial structure of the dielectric decrement}

We have so far analyzed the salt-induced decrement of the bulk permittivity~(\ref{eq55}) characterizing the dielectric screening ability of the electrolyte at large distances from charge sources. Here, we scrutinize the spatial structure of the dielectric decrement in the close vicinity of charges. 

\subsubsection{Modification of the dielectric spectrum by added salt}

The electrolyte permittivity can be experimentally measured at different wavelengths~\cite{Hasted1948,Haggis1952,Harris1957,Hasted1958}. Motivated by this point, we characterize first the spatial spectrum of the salt-dressed permittivity. To this aim, in Figs.~\ref{fig4}(a)-(b), we reported the Fourier-transformed electrolyte permittivity~(\ref{eq46III}) and its salt component~(\ref{eq48}) at various solute concentrations. One sees that as the wavelength $q$ rises, the pure solvent permittivity (black) uniformly drops from the bulk permittivity $\e_{\rm w,b}$ to the vacuum permittivity  $\te\w(q\to\infty)=1$ corresponding to the dielectric void in the vicinity of charge sources~\cite{PRE1,JCP2013}. However, upon salt addition, the dielectric spectrum acquires a non-uniform dependence on the wavelength.

To understand the change of the dielectric spectrum, we note that in Fig.~\ref{fig4}(b), salt addition into the solvent gives rise to a substantial dielectric decrement ($\delta\te(q)<0$) at short wavelengths $qa\lesssim2$ and a weak dielectric increment ($\delta\te(q)>0$) in the regime $qa\gtrsim2$ of long wavelengths. As a result,  Fig.~\ref{fig4}(a) shows that added salt lowers the dielectric permittivity in the short wavelength regime and moves the peak of the permittivity spectrum from the strict IR limit $q=0$ to a finite wavelength value $q=q^*$ rising with the salt concentration ($\rho_{\rm s}\uparrow q^*\uparrow$). Next, we investigate the impact of this peculiarity on the range of the non-local dielectric screening in real space.

\subsubsection{Effect of added salt on the spatial permittivity profile}

Figs.~\ref{fig4}(c)-(d) display the permittivity profile in Eq.~(\ref{elpr}) and the salt-driven decrement function~(\ref{delpr}) at various solute concentrations. In the case of a pure solvent (solid black curve), the permittivity is characterized by a region of dielectric deficiency ($\e(r)<\e_{\rm w,b}$) followed by a high dielectric increment peak ($\e(r)\gg\e_{\rm w,b}$) and an oscillatory convergence towards the bulk permittivity $\e_{\rm w,b}$ (dotted curve). In Refs.~\cite{PRE1,JCP2013}, we showed that the corresponding dielectric structure equally observed in AFM experiments~\cite{expdiel} and MD simulations~\cite{Hans,prlnetz} originates from the particularly strong self-screening of the polarization charges at the large solvent concentration.

In Figs.~\ref{fig4}(c)-(d), one sees that salt addition into the solvent leads to the spatially non-uniform decrement of the liquid permittivity; one notes that salt-solvent correlations lead to the maximum amount of dielectric decrement  around the first permittivity peak where the dielectric response of the solvent is the strongest, while the liquid region around $r\approx a$ charactarized by the weakest dielectric response experiences an insignificant dielectric decrement. One also sees that added salt approaches the location of the dielectric peak from $r\approx a/2$ towards the source charge located at $r=0$. This indicates that the addition of salt reduces not only the magnitude but also the range of the dielectric screening by solvent.

For a systematic characterization of the correlation between the strength of the solvent-solute interactions and the liquid permittivity profile, in Figs.~\ref{fig4}(e)-(f), we plotted the dielectric permittivity and the decrement functions at various HC diameters. The plots show that as the HC diameter is lowered from $d=1.5\;a$ to $a$, the intensification of the salt-solvent correlations flattens the  permittivity profile by suppressing the minima and maxima beyond the first dielectric peak. Hence, in addition to lowering the range and the magnitude of the liquid permittivity, salt-solvent correlations also attenuate the dielectric structure in the vicinity of charge sources and suppress the non-locality of the electrostatic interactions.

\section{Conclusions}

Within a field-theoretic explicit solvent formalism, we characterized the electrostatic mechanism behind the salt-driven dielectric decrement. By accounting for the finite size of the solvent molecules, we avoided the UV divergence issue bypassed in point-dipole models with a UV cutoff. The key novelty of our formalism is the relaxation of the WC treatment of salt used by earlier models of dipolar liquids~\cite{Andelman2013,Andelman2018,JCP2013}. More precisely, by exploiting the contrast between the concentration of the dense solvent and the substantially more dilute salt component of the electrolyte, we incorporated the salt fluctuations via a virial expansion. Upon this virial approximation that enabled the explicit inclusion of the many-body salt-solvent interactions, the experimentally observed linear dielectric decrement by dilute salt addition emerged directly in the form of the permittivity formula~(\ref{eq55}).

The analytical formula~(\ref{eq55}) allows to identify clearly the electrostatic mechanism behind the dielectric decrement as the salt screening of the polarization charges suppressing the dielectric response of the polar liquid. The comparison of the theoretical prediction~(\ref{eq55}) and the experimental permittivity data indicates that the virial approximation underlying our formalism holds up to the characteristic salt concentration $\rho_{\rm s}\sim2$ M.  It is noteworthy that the latter is much larger than the characteristic multivalent ion density $\rho_{\rm s}\sim10^{-3}$ M marking the validity regime of the solvent-implicit dressed ion approaches employing a similar virial approximation~\cite{Podgornik2011,JPCB2020,Langmuir2022,JCP2020}. This contrast can be explained by the fact that the dominant component of solvent-implicit electrolyte mixtures is the monovalent salt of submolar concentration, while the major component of our explicit solvent model is the polar liquid of substantially larger density $\rho\w=55$ M.

By direct confrontation with experimental data, we showed that our theory can reproduce qualitatively the decrease of the electrolyte permittivity with the rise of the temperature, as well as the thermal decay of the dielectric decrement with reasonable quantitative accuracy. Considering that our formalism models salt ions as point-charges and therefore neglects the thermal variation of the electronic cloud radius determining the HC size, the overall agreement with experiments reached with a single adjusted value of the HC diameter is rather remarkable. Via the incorporation of the ionic polarizability into our model, the point charge approximation for the salt ions can be relaxed in future works. This extension would equally allow to include the ion specificity into our model.

It was shown that that the attenuation of the dielectric decrement by the rise of the temperature originates from the suppression of the solvent-ion correlations by thermal fluctuations. We also found that due to the amplification of the same correlations with rising valency, added 2:1 salts induce a more efficient dielectric decrement than 1:1 salts. This prediction agrees with the experiments of Harris and O'Konski carried out with symmetric and asymmetric electrolytes~\cite{Harris1957}.

The inspection of Fig.~\ref{fig3} shows that Eq.~(\ref{eq55}) predicts the thermal variation of the reduced permittivity $\Delta\e_{\rm b}$ more accurately than the net electrolyte permittivity $\e_{\rm b}$ dominated by the background solvent. This point indicates that in Fig.~\ref{fig2}(b), the underestimation of the thermal susceptibility of the permittivity data by our formalism may originate from the gaussian-level treatment of the pure solvent correlations leading to the Langevin identity~(\ref{lang}). Moreover, in Fig.~\ref{fig2}(c), one sees that at large HC radii, the dielectric decrement is followed by a weak increment regime where the electrolyte permittivity slightly exceeds the solvent permittivity. While we could not ascertain the reason behind this trend, the effect might be again an artefact of the WC treatment of the dipole-dipole correlations unable to account for the incompressibility of water. In a future work, we plan to consider the effect of the incompressible liquid condition as well as the excluded-volume associated with the salt and solvent molecules. We finally note that the predictions of the MD simulations with higher complexity level equally exhibit quantitative disagreement with the experimental permittivity data, the deviation exceeding $20\%$ for some force field models~\cite{Saric2020}. This indicates that the quantitatively accurate characterization of the dielectric response of pure water remains an open challenge.

\smallskip
\textbf{Conflicts of interest}

There are no conflicts to declare.

\smallskip
\appendix
\section{Derivation of the electrostatic SD equations}
\label{ap1}

In this appendix, we review the derivation of the electrostatic SD identities introduced in Ref.~\cite{JCP2020}. For this purpose, we define first the following functional integral
\be\label{APeq20III}
I=\int\mathcal{D}\ph\;e^{-\beta H_{\rm el}[\ph]}F[\ph].
\ee 
Then, we introduce in Eq.~(\ref{APeq20III}) the infinitesimal shift of the electrostatic potential $\phi\ce(\br)\to\phi\ce(\br)+\delta\phi\ce(\br)$, and linearize the result in terms of the function $\delta\phi\ce(\br)$. The resulting variation of the integral~(\ref{APeq20III}) follows as
\bea\label{APeq20IV}
\delta I&=&\int\mathrm{d}\br\delta\phi\ce(\br)\int\mathcal{D}\ph e^{-\beta H_{\rm el}[\ph]}\\
&&\hspace{2.3cm}\times\left\{\frac{\delta F[\ph]}{\delta\phi\ce(\br)}-F[\ph]\frac{\delta H_{\rm el}[\ph]}{\delta\phi\ce(\br)}\right\}.\nonumber
\eea
At this point, we note that the potential shift $\delta\phi\ce(\br)$ can be removed via the redefinition of the integration measure in Eq.~(\ref{APeq20III}), implying the invariance of the integral~(\ref{APeq20III}) under the shift by $\delta\phi\ce(\br)$, i.e. $\delta I=0$. Thus, setting the r.h.s. of Eq.~(\ref{APeq20IV}) to zero, and dividing the result by the partition function~(\ref{eq12}), one gets
\be
\label{APeqG}
\lan\frac{\delta F[\ph]}{\delta\phi\ce(\br)}\ran=\lan F[\ph]\frac{\delta H_{\rm el}[\ph]}{\delta\phi\ce(\br)}\ran.
\ee
By setting in Eq.~(\ref{APeqG}) $F[\ph]=1$ and $F[\ph]=\phi\ce(\br')$, one obtains respectively the formally exact SD equations
\bea\label{APSD1}
&&\lan\frac{\delta\left(\beta H_{\rm el}[\ph]\right)}{\delta\phi\ce(\br)}\ran=0,\\
\label{APSD2}
&&\lan\frac{\delta\left(\beta H_{\rm el}[\ph]\right)}{\delta\phi\ce(\br)}\phi\ce(\br')\ran=\delta(\br-\br').
\eea


\begin{thebibliography}{99}
\bibitem{rem1} This numerically revealed feature that we have not been able to explain by a simple mathematical argument implies that the second term in the curly bracket of Eq.~(\ref{eq54}) exactly cancels out the first term proportional to the susceptibility of the pure solvent.
\end{thebibliography}
\end{document}